\documentclass{sig-alternate}

\usepackage[utf8]{inputenc}

\usepackage{booktabs}
\usepackage{caption}
\usepackage{subcaption}
\usepackage{float}
\usepackage{multirow}
\usepackage{url}

\usepackage{url}
\makeatletter %
\g@addto@macro{\UrlBreaks}{\UrlOrds} %
\makeatother %

\begin{document}


\title{Characterizing videos, audience and advertising in Youtube channels for kids}

\numberofauthors{6}

\author{
\alignauthor
Camila Souza Ara{\'u}jo $^1$ $^\star$
\email{\small camilaaraujo@dcc.ufmg.br}
\alignauthor
Gabriel Magno  $^1$ $^\star$
\email{\small magno@dcc.ufmg.br}
\alignauthor
Wagner Meira Jr. $^1$
\email{\small meira@dcc.ufmg.br}
\and
\alignauthor
Virgilio Almeida $^{1,2}$
\email{\small valmeida@cyber.law.harvard.edu}
\alignauthor
Pedro Hartung $^3$
\email{\small pduartehartung@mail.law.harvard.edu}
\alignauthor
Danilo Doneda $^4$
\email{\small danilo@doneda.net}
\and
\affaddr{$^1$ Universidade Federal de Minas Gerais, Belo Horizonte, Brazil}  \\
\affaddr{$^2$ Berkman Klein Center, Harvard University, Cambridge, USA}\\
\affaddr{$^3$ Harvard Law School, Cambridge, USA}\\
\affaddr{$^4$ Universidade do Estado do Rio de Janeiro, Rio de Janeiro, Brazil}\\
\affaddr{$^\star$ \small{These authors contributed equally to this work}}\\
}


\maketitle

\begin{abstract}
Online video services, messaging systems, games  and social media services are tremendously popular among young people and children in many countries. Most of the digital services offered on the internet are advertising funded, which makes advertising ubiquitous in children’s everyday life. To understand the impact of advertising-based  digital services on children, we study the collective behavior of users of YouTube for kids channels and present  the  demographics of a large number of users. We collected data from 12,848 videos from 17 channels in US and UK and 24 channels in Brazil. The channels in English have been viewed more than 37 billion times. We also collected more than 14 million comments made by users. Based on a combination of text-analysis  and  face recognition tools, we  show the presence of racial and gender biases in our large sample of users. We also identify children actively using YouTube, although the minimum age for using the service is 13 years in most countries. We provide comparisons of user behavior among the three countries, which represent large user populations in the global North and the global South.
\end{abstract}

\section{Introduction}

All over the world, digital technologies are shaping children’s lives for better or for worse. Policy makers, researchers and educators who work with children’s rights agenda recognize the social impact of digitization for young people’s lives \cite{jeff2015how, livingstone2014children,  livingstone2014digital, Dehghani2016165, Keeffe800}. Online video services (e.g., YouTube, Netflix, BBC, etc.), messaging systems (e.g., Whatsapp, Messenger, etc.), games (e.g., Apple, Google Play, IGN, Gamespot, etc.) and social media services (e.g., Snapchat, Facebook, etc.) are tremendously popular among young people in many countries~\cite{lupianez2016study}. YouTube in particular has been viewed as an alternative to traditional children’s TV\cite{dredge2015why}. Millions of children are already watching videos on YouTube, most of them logged in from their parents’ accounts. For example, the channel of a popular youtuber, Joseph Garrett,  has 7.8 million subscribers and its videos have been viewed 5.3 billion times, making it one of the most popular British YouTube channels for children~\cite{Guardian2016}. Most of the digital services offered in the Web are funded by advertising, which makes advertising ubiquitous in children’s everyday life. YouTube offers ad-funded video channels. As a consequence, several questions about the role of advertising in children’s life arise. What forms of advertising do children face on the internet? How do children react to online advertising?  The goal of this paper is to provide a detailed quantitative characterization of users, videos and advertising in a sample based on several popular YouTube channels for kids in US, UK and Brazil.

There are several types of ads. For example, advergames are video games created by a company with the intention of promoting the company itself or its products. Usually these games are distributed freely as a marketing tool. There are cases of food and drink companies that target children with advertising unhealthy products on various internet platforms. An European Commission study~\cite{lupianez2016study} reports that online marketing to children and young people is widespread, and in some cases various marketing techniques used are not always transparent to the child consumer. There are marketing strategies that target children on YouTube with advertising disguised as other content. They use popular youtubers to pitch products and brands as non-commercial content in videos that are viewed worldwide. Characterizing and understanding these strategies and their effectiveness is a key task for making the internet and the web a better place for children. 

Recent figures published by ITU (International Telecommunication Union) in 2016 show that developing countries now account for the vast majority of internet users, with 2.5 billion users compared to one billion in developed countries. According to~\cite{global2016one}, one of every three internet users in the world is a child. internet is becoming the main medium through which children collaborate, share, learn and play. In order to understand rights, risks and opportunities for children on the internet, it is important to look at countries from both the global North and the global South~\cite{livingstone2014digital}. Because of its worldwide penetration, YouTube channels for kids is a good scenario to understand advertising campaigns that target children.  The paper provides a study of interaction of online advertising and Youtube for Kids audience in US, UK and Brazil. The survey on digital marketing by the company GroupM\footnote{www.groupm.com} in April 2017, estimates 44 million YouTube users in UK and 72 million in Brazil. The Statistics Portal\footnote{www.statista.com} estimates 180 million YouTube users in US. 

In order to collect and analyze YouTube usage data, we developed an experimental methodology based on the combination of free APIs and open source tools available on the internet. The results of the  characterization presented in this paper can be useful for policy makers in different countries to assess the need of public policies to protect children online. To the best of our knowledge, this paper is the first one to study and characterize videos, audience and advertising in internet channels for children. 

Overall, we make  the following  contributions: 
\begin{enumerate}
\item We develop a simple experimental methodology to collect and analyze large amounts of  YouTube usage data based on APIs and open source tools available on the internet. 
\item We integrate  free text-analysis and face recognition tools   to identify age, race and gender of YouTube channel users  as well as to characterize the behavior of those users. 
\item We identify children actively using YouTube, although the minimum age for using the platform is 13 years, according to their Terms of Service. Even if some usage of under 13 is generally considered as fair due to parents’ or legal responsible consent and supervision, the fact is that if children are actually using the platform they can be exposed to advertising, what raise concerns about compliance with publicity regulation in several countries.
\item We show  the presence of racial and gender bias   in the large  sample of  YouTube  users in our data sets.  The percentage of black users is very small when compared to white and Asian users. 
\item We analyze  the behavior of YouTube users in three countries, US and UK in the global North and Brazil in the global South. We show  differences and similarities in the demographics of YouTube channel  audience as well as in the categories of products and brands associated with the videos of the channels.
\end{enumerate}
The rest of the paper is organized as follows. We begin with a description of research questions associated with online advertising for children in section~\ref{sec:research}. In section~\ref{sec:computational}, we discuss the computational approach used to gather and analyze data from different internet channels for kids. A detailed description of the datasets collected from YouTube channels is given in section~\ref{sec:dataset}. Next, in sections \ref{sec:ana1} and \ref{sec:ana2}, we characterize videos and advertising of popular YouTube channels. Finally we describe and characterize the behaviors of users of YouTube channels in section \ref{sec:ana3}. Section \ref{conclusion} summarizes our findings and discusses future work.

\section{Research Questions}
\label{sec:research}

In this section we discuss the research questions that we address  in our work.
The expansion of the use of social and digital media led to the expansion of the presence of marketing to children through digital platforms. As mentioned, advergames, product placement in YouTube videos and online games, marketing in social networks and other strategies are commonly used by companies to attract the attention of children and persuade them to consume certain products or services. However, unlike traditional media, marketing in the digital environment takes new forms and many of them are more difficult to be clearly identified.

By providing a detailed characterization of YouTube channels for kids, this paper aims at shedding  some light on streaming video-on-demand programming that target children all over the world.  It also seeks to understand how the children's audience interacts with channels and videos through the children's engagement in the conversations in the video comments in YouTube\cite{jeff2015how}.  As a consequence of our research goal, we ask the following questions: 
\begin{itemize}
\item What are the characteristics of the content of the most recurrent videos on children’s channels?
\item What does characterize the audience to the videos for children (e.g., is there a predominance of gender in the audience and also in the young youtubers)?
\item Which classes of  products are marketed to a specific target (i.e., gender, age, ethnicity)?  
\item Is it possible to measure the percentage of children’s audience in the YouTube channels examined? 
\item What are the gender, specific age and ethnicity among the children identified? 
\item What is the content of the most recurrent videos on children's channels?
\item Is it possible to identify publicity directly aimed at children on the channels?
\end{itemize}
In order to investigate the stated research questions, this study relies on  data collected from popular YouTube channels in US, UK and Brazil. We developed a computational methodology based on open source code to analyze the data.

\section{Related Work}

\begin{figure*}[!h]
\centerline{
\includegraphics[width=14cm]{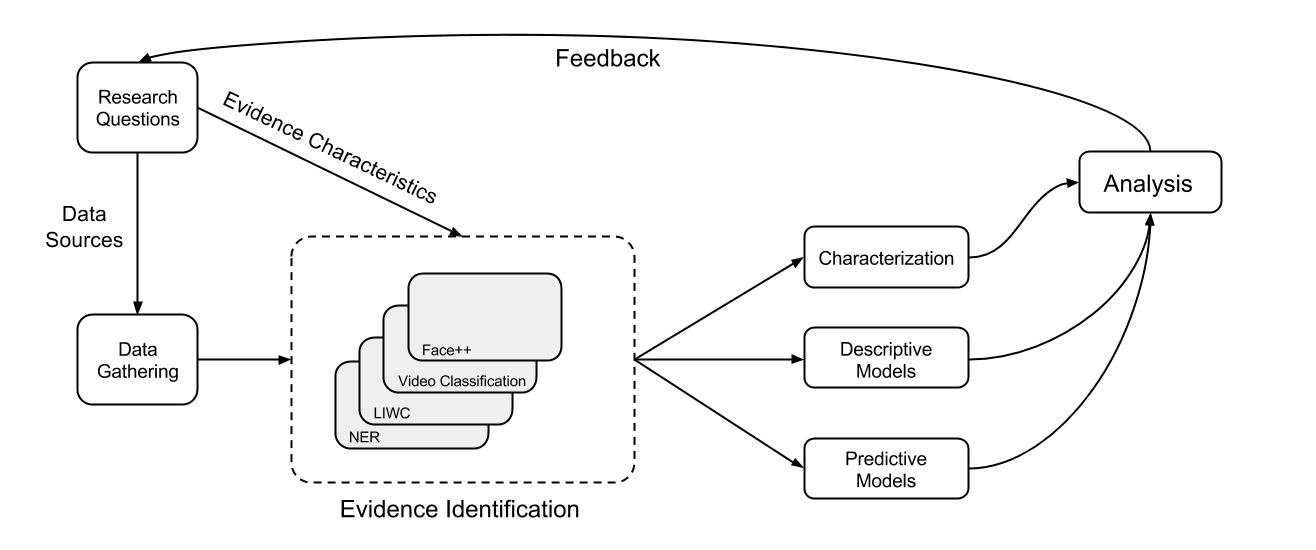}}
\caption{Methodology}
\label{fig:methodology}
\end{figure*}

We now briefly summarize existing studies related to YouTube analysis, as well as studies of user behavior in Social Media. 

Online social networks are popular platforms for people to connect and interact with each other~\cite{raphael2013ladies}. According to Benevenuto et al.~\cite{benevenuto2009characterizing}, understanding users behavior on social networking sites creates several opportunities. For example, accurate models of user behavior are important in social studies and viral marketing, since viral marketers may exploit models of user interaction to spread their content quickly and widely. Nowadays, many children use the Internet and mobile technologies as part of their everyday lives. The overlap of the online and offline world comes with a range of digitally-mediated opportunities and risks. Reference \cite{Keeffe800} provides a qualitative analysis of different social media sites to assess if they provide healthy environments for children and teenagers. In \cite{Przybylski2017}, the authors investigate the effectiveness of Internet filtering tools designed to shield teenagers from aversive online experiences. Based on $1,030$ in-home interviews conducted with early teenagers aged from 12 to 15 years, the paper shows that Internet filters were not effective at shielding early teenagers from aversive online experiences, that include scary online videos. Magno et al.~\cite{magno2012new} studied the Google+ environment and compared its network structure with Facebook and Twitter, and noticed that it has a higher average path length and higher reciprocity. They also compare the user profile characteristics between different countries, and found that some countries are more private than others.

As pointed out by Benevenuto et al.~\cite{benevenuto2010characterization}, online video sharing systems have been increasing and gaining popularity. These environments allow several kinds of interactions between users and videos, such as publication of comments. This is the most related work to ours in terms of characterization methodologies, however the authors presented an in-depth workload characterization of sessions and requests on an video server different than YouTube. In terms of sociological studies of the impact of YouTube on kids, according to the authors of \cite{stoilova2016global}, the evidence on how use of the Internet impacts on child rights and well-being is still scattered and patchy in most countries. Livingstone and Local \cite{livingstone2017measurement} discussed the problem of audience measurement techniques regarding children’s television viewing, because of the diversification in devices on which television content can be viewed. It is already well known that television content can be viewed on Internet-enabled devices and Internet content can be accessed via Internet-enabled television sets, but such viewing cannot be measured satisfactorily at present.

There are few articles in the literature that address quantitative analysis of online advertising for children in Internet video channels. For example, Dehghani et al.~\cite{Dehghani2016165} uses data collected  from Italian students to analyze the perception of YouTube by young people. Data were obtained from 315 questionnaires. The results show positive aspects of YouTube in terms of entertainment, informativeness and customization. The negative aspect is related to YouTube advertising. Unlike this reference, our paper relies on a large scale datasets collected from YouTube channels.

\section{Experimental Methodology}
\label{sec:computational}

In this section we describe the computational approach adopted to answer the aforementioned research questions.

\subsection{Rationale}

The research questions outlined in Section~\ref{sec:research} are  hard to verify and and quantify, mainly considering that most of our input data  that are publicly available and are composed of free text and images. One immediate consequence is that it is unfeasible, in practice, to get fully accurate and complete datasets about the video body we want to analyze. Then, we adopt an approach based on identifying evidences. Each evidence makes explicit a piece of information about the entity being analyzed. Considering YouTube users, examples of evidences are his or her gender, age and race, as extracted from a profile picture. In this work, we chose a set of evidences that demonstrate the occurrence of advertising in child-oriented videos, as described in Section~\ref{sec:methodology}. 

These evidences should be conservative, although it is possible to improve the gathering techniques and be able to identify evidences. For instance, when we label a video as an advertising piece, we should be as sure as possible that it really is. The immediate impact of our approach is that all our figures are lower bounds of the actual evidence counts. Although we are usually not able to perform analyses that demand accurate counts, they demonstrate clearly the occurrence of targeted phenomenon or behavior. It is important to emphasize that our strategy also leverages on the fact that there are already a huge number of techniques and tools that may be promptly used for identifying evidence, as we discuss next.

\subsection{Methodology}
\label{sec:methodology}

In this section we present our methodology for assessing the occurrence of not only advertising in YouTube videos, but may also serve to analyze the occurrence of various phenomena associated with Internet applications. As we discussed in Section~\ref{sec:research}, we look for identifying and modeling three groups of evidences: (i) content characterization; (ii) audience profiles; and (iii) detection of products and their publicity in videos.

The starting point of our methodology is the set of research questions we want to answer. 
We then select the data sources from which we will gather data for each question. 
We also map the questions into evidences to be identified, which demand the application of one or more techniques to the data, usually enhancing them. Examples of evidence employed in this paper are the positivity of video meta-data and user profile inferred based on face snapshots, which add attributes to both video and users, respectively. The evidences may be characterization findings, descriptive models or predictive models. Characterization may use summary statistics, among other techniques, to detect invariants, trends and other properties. Descriptive models comprise patterns and models inherent to the data, such as clusters and correlations. Predictive models estimate samples' class or numerical dependent variables. It is worth mentioning that there is a large spectrum of techniques for identifying and modeling evidences, most of them freely available in the Internet. The enriched data, characterization findings and derived models are then used for analysis and answering of the original research questions. The last step is to, considering the correctness and completeness of the answers, improve the whole process towards increasing its quality. Figure~\ref{fig:methodology} depicts the methodology proposed.

In the next sections we discuss the techniques used for evidence identification and modeling in more detail, as well as how they help our analysis.

\subsection{Data Gathering}

YouTube is a large-scale video sharing online platform where users can produce and/or consume content. On YouTube users need first to create a channel to upload videos. Users do not need to be logged in to watch videos, but they need to be logged in to comment and 'like' videos. We collect information of YouTube videos and comments using Google's YouTube Data API~\footnote{\url{https://developers.google.com/youtube/v3/}}, accessing it directly with Python 3 scripts. The data collection process was performed in 5 phases:

\begin{enumerate}
\item \textbf{Select list of channels}: we manually select a list of 41 popular YouTube channels targeted to children. This selection was performed by children's rights experts and based on their popularity and also empirical evidence they may employ advertising strategies.
\item \textbf{Retrieve list of videos}: for each channel, we collect its list of videos. Due to API limits, we only get the last 500 videos published in the channel.
\item \textbf{Retrieve video statistics}: for each video, we collect its information and statistics.
\item \textbf{Retrieve comments}: for each video, we gather the comments published by users about it.
\item \textbf{Retrieve replies}: YouTube users may reply to a video comment, thus for each comment we collect its list of replies as well. In our analysis we handle replies as normal comments.
\end{enumerate}

As we discuss later, the collected information allows extensive analysis about the video characteristics, the marketing strategies it may employ, and the observed impact of such strategies.

\subsection{Evidence Identification and Modeling}

In this section we present the various strategies and techniques used in this study for sake of evidence identification and modeling.

\subsubsection{Entity Recognition}
\label{sec:entityrec}

In order to characterize the videos we need to extract entities (names, brands, products etc.) that are mentioned in it. We use a technique called Named Entity Recognition (NER), a method that labels sequences of words into categories of things, such as company names, person and cities. We use the Stanford NER~\footnote{\url{https://nlp.stanford.edu/software/CRF-NER.shtml}} tool with the English pre-trained model. Unfortunately, it does not have a pre-trained model for Portuguese, so we use this technique only for the videos of U.S channels. For both Portuguese and English, we also detected entities, in particular products and brands, by assessing the video meta-data, as discussed in Section~\ref{sec:method_video_classification}.

\subsubsection{Sentiment Analysis}

We assess the public perception on the videos published by analyzing the content of the comments written about them. To attain this task, we use the Language Inquiry and Word Count (LIWC)~\cite{paper_liwc2007}, a lexicon used to verify the occurrence of words from several grammatical (e.g., pronouns, verbs, and articles, among others) and semantic (e.g., positive emotion, social, motion) categories. We use the LIWC 2007 dictionary, whose complete list of categories and word examples is available at LIWC's website~\footnote{\url{https://liwc.wpengine.com/}}. For sake of our analysis in this work, we calculate the proportion of comments that contained at least one word of a particular category and label the comment to all categories that match.

\subsubsection{Product Category Identification}
\label{sec:method_video_classification}

A key component for our analysis is the identification of products and respective products categories that may be marketed and advertised in the videos being analyzed.
However, watching and labeling thousands or millions of videos is unfeasible. We adopt the strategy described next for identifying product categories present in each video. We start by extracting the tags of the "video tags" field for each video, which is a list of labels manually inserted by the owner of the channel. For the videos of U.S channels, we also consider the entities mentioned in the "description" field, which are extracted using the NER tool (as described in Section~\ref{sec:entityrec}). 

The next step is, given the list of tags (manual tags and NER tags) for each video, we calculate the frequency of the tags among all videos, and compile a sorted list of the most popular tags for each country. Then, we retrieve the top 1,000 tags in each list and manually check the assignment of each tag to one of the 23 categories of products presented in Table~\ref{tab:video_categories}. If the tag does not match any of the categories, we ignore it.

\begin{table}[!ht]
\centering
\caption{Categories of Products Used to Classify Videos}
\label{tab:video_categories}
\begin{tabular}{@{}lll@{}}
\toprule
\multicolumn{3}{c}{Product Categories}                                                     \\ \midrule
\multicolumn{1}{l|}{Footwear}  & \multicolumn{1}{l|}{Water}        & School Suplies        \\
\multicolumn{1}{l|}{Clothes}   & \multicolumn{1}{l|}{Snacks}       & Electronics           \\
\multicolumn{1}{l|}{Fast Food} & \multicolumn{1}{l|}{Fresh Food}   & Pet Products          \\
\multicolumn{1}{l|}{Chocolate} & \multicolumn{1}{l|}{Food (other)} & Books                 \\
\multicolumn{1}{l|}{Candies}   & \multicolumn{1}{l|}{Cosmetics}    & Travel and Recreation \\
\multicolumn{1}{l|}{Sodas}     & \multicolumn{1}{l|}{Make Up}      & Services              \\
\multicolumn{1}{l|}{Juices}    & \multicolumn{1}{l|}{Toys}         & Movies and Shows      \\
\multicolumn{1}{l|}{Yogurts}   & \multicolumn{1}{l|}{Games}        &                       \\ \bottomrule
\end{tabular}
\end{table}

\subsubsection{Video Classification}

Once we assigned tags to product categories, we may classify the videos by simply verifying whether it contains one of the tags of a particular category. It is important to notice that a tag might have been classified into more than one category (e.g "Disney toys" is both from "Toys" and "Movies and Shows" category). In the same sense, a video might be assigned to two or more categories.  Notice that the process we employed provides good precision, but not necessarily a good recall, since it relies on the channel owner, who provides the tag definition and description information. It is beyond the scope of this work to assess how complete, accurate, and consistent across videos and channels these data are.

\subsubsection{User Visual Profiling}

The user visual profiling aims to determine user information such as gender, age and race of the users who comment the videos. As previously stated, after selecting a channel list, we collect information from the last 500 videos of each channel, which includes the URL of the YouTube profile image of all users who left comments. We then download the profile pictures associated with all users and use Face++\footnote{https://www.faceplusplus.com/} to extract information such as age, race and gender about each face in the photo. Face++ is an online API for facial recognition and its accuracy is known to be over 90\% for face detection \cite{bakhshi2014faces}. It is important to note that not all users use real photos as a profile image, so Face++ is not always able to identify a face. In the next section, we will present the number of identified faces that composes our dataset. It is important to mention that, although we employed just visual profiling, any technique that provides such information may be used. The key issue here is the coverage of the profiling information acquired considering the user population and their accuracy.

\section{Data Analysis and Results}
\subsection{Datasets}
\label{sec:dataset}

\begin{figure}[!ht]
\centering
\includegraphics[scale=0.17]{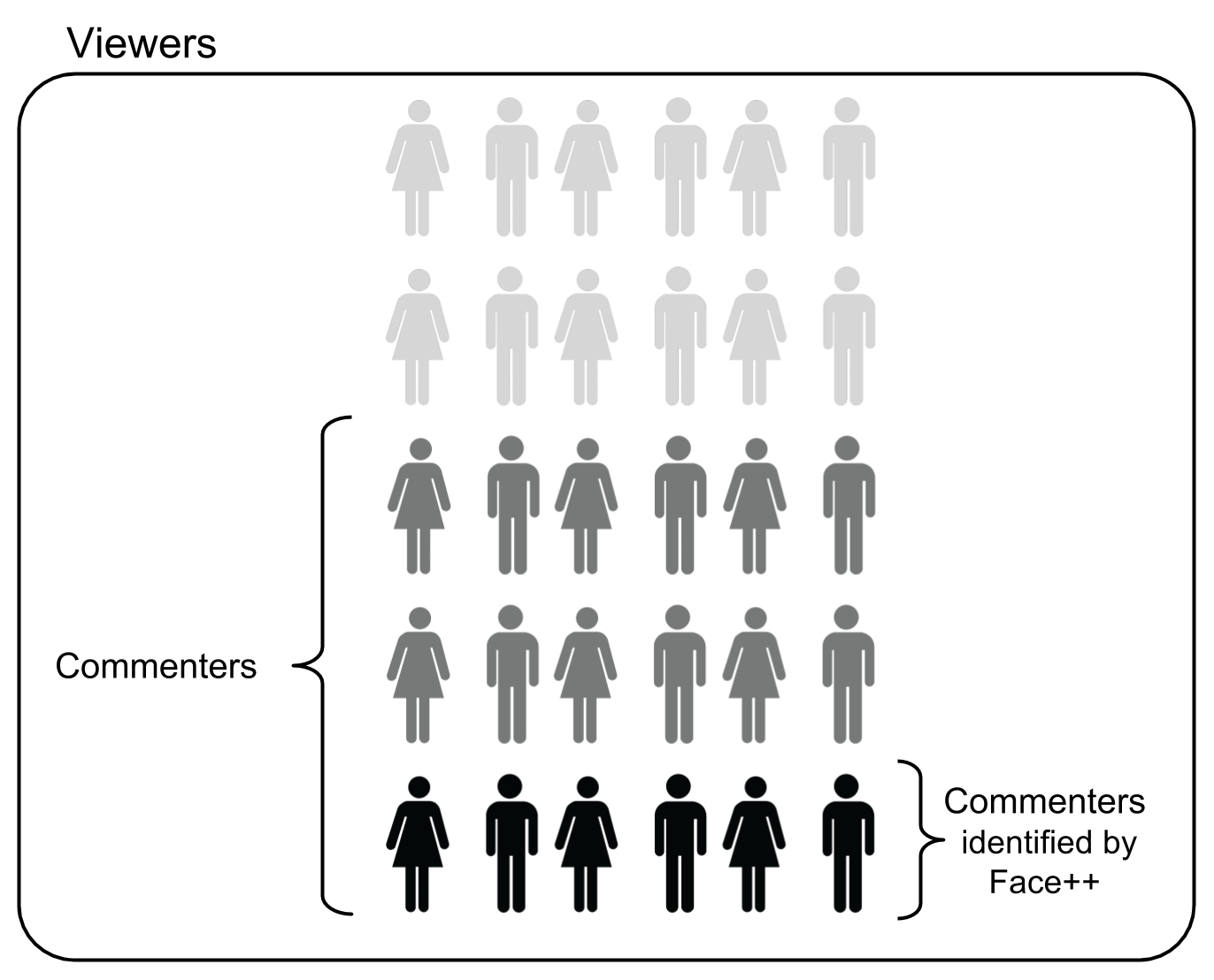}
\caption{'Viewers' are all users who watched the video and 'commenters' are users who left comments. We still have a subset of commenters for whom Face++ has identified a face.} 
\label{fig:users}
\end{figure}

Now we present a brief characterization of the datasets collected for this work. We chose to collect data from 24 Brazilian YouTube channels, and 17 YouTube channels for kids produced in English from United States and United Kingdom. The rationale for selecting US, UK and Brazil is the following. Most of the YouTube content is produced in English\footnote{https://medium.com/@synopsi/what-youtube-looks-like-in-a-day-infographic-d23f8156e599}. However,   Brazil is the second largest market considering time spent on YouTube\footnote{https://googlediscovery.com/2017/03/23/google-brasil-dados-importantes-sobre-o-google-no-brasil/}. The three countries represent a large number of YouTube users in the global North and global South. Our Brazilian dataset comprises data about 7,664 videos and 10,940,565 comments associated with them, issued by 2,982,595 distinct users. That is, the same user may leave more than one comment. It is important to emphasize that throughout the work when we refer to 'users' we do not refer to all users who watched the video, we refer to the subset of users who made comments, as shown in the Figure~\ref{fig:users}. From now on, we will call those users - users who left comments - of 'commenters', in this way we avoid confusing them with the users who only watch the videos. Of this total of commenters we were able to extract information of 129,286 faces. Table~\ref{tab:datasetsum} summarizes the size of the dataset collected from the channels of both countries. In Table~\ref{tab:datasetfields} we summarize the dataset composition.

\begin{table}[!ht]
\centering
\caption{Dataset Summary}
\label{tab:datasetsum}
\begin{scriptsize}
\begin{tabular}{lccr}
\cline{2-4}
       & \multicolumn{1}{l}{\textbf{\#channels}} & \textbf{\#videos}             & \multicolumn{1}{c}{\textbf{\#views}} \\ \hline
Brazil & 24                                      & 7,664                         & 4,614,161,928                        \\
US+UK  & 17                                      & 5,184                         & 37,401,690,211                       \\ \hline
       & \multicolumn{1}{l}{}                    & \multicolumn{1}{l}{}          & \multicolumn{1}{l}{}                 \\ \cline{2-4} 
       & \textbf{\#comments}                     & \textbf{\#commenters}         & \multicolumn{1}{c}{\textbf{\#faces}} \\ \hline
Brazil & \multicolumn{1}{r}{10,940,565}          & \multicolumn{1}{r}{2,982,595} & 129,286                              \\
US+UK  & \multicolumn{1}{r}{3,569,553}           & \multicolumn{1}{r}{2,013,419} & 9,248                                \\ \hline
\end{tabular}
\end{scriptsize}
\end{table}

\begin{table}[!h]
\centering
\caption{Dataset Fields}
\label{tab:datasetfields}
\begin{scriptsize}
\begin{tabular}{ccl}
\hline
\multicolumn{3}{c}{\textbf{Dataset Fields}}                                                                                                                                                                                                         \\ \hline
\multicolumn{1}{c|}{\multirow{13}{*}{\textit{Video}}}   & \multicolumn{1}{c|}{channel id}                                                       & Video id                                                                                          \\ \cline{2-3} 
\multicolumn{1}{c|}{}                                   & \multicolumn{1}{c|}{channel name}                                                     & \begin{tabular}[c]{@{}l@{}}Id of channel where video was \\ posted\end{tabular}                   \\ \cline{2-3} 
\multicolumn{1}{c|}{}                                   & \multicolumn{1}{c|}{video id}                                                         & Video channel name                                                                                \\ \cline{2-3} 
\multicolumn{1}{c|}{}                                   & \multicolumn{1}{c|}{video title}                                                      & Title of the video                                                                                \\ \cline{2-3} 
\multicolumn{1}{c|}{}                                   & \multicolumn{1}{c|}{video description}                                                & \begin{tabular}[c]{@{}l@{}}Video description (made \\ manually by youtuber)\end{tabular}          \\ \cline{2-3} 
\multicolumn{1}{c|}{}                                   & \multicolumn{1}{c|}{transcript}                                                       & \begin{tabular}[c]{@{}l@{}}Automatic textual transcription \\ from the audio\end{tabular}         \\ \cline{2-3} 
\multicolumn{1}{c|}{}                                   & \multicolumn{1}{c|}{subtitle}                                                         & \begin{tabular}[c]{@{}l@{}}Manual subtitle (made by \\ youtuber or by third parties)\end{tabular} \\ \cline{2-3} 
\multicolumn{1}{c|}{}                                   & \multicolumn{1}{c|}{video tags}                                                       & \begin{tabular}[c]{@{}l@{}}List of tags (made manually \\ by youtuber)\end{tabular}               \\ \cline{2-3} 
\multicolumn{1}{c|}{}                                   & \multicolumn{1}{c|}{video date}                                                       & Video posting date and time                                                                       \\ \cline{2-3} 
\multicolumn{1}{c|}{}                                   & \multicolumn{1}{c|}{video duration}                                                   & Video duration in seconds                                                                         \\ \cline{2-3} 
\multicolumn{1}{c|}{}                                   & \multicolumn{1}{c|}{view count}                                                       & Number of views                                                                                   \\ \cline{2-3} 
\multicolumn{1}{c|}{}                                   & \multicolumn{1}{c|}{comment count}                                                    & Number of comments                                                                                \\ \cline{2-3} 
\multicolumn{1}{c|}{}                                   & \multicolumn{1}{c|}{like count}                                                       & Number of likes                                                                                   \\ \hline
\multicolumn{1}{c|}{\multirow{10}{*}{\textit{Comment}}} & \multicolumn{1}{c|}{comment id}                                                       & Comment id                                                                                        \\ \cline{2-3} 
\multicolumn{1}{c|}{}                                   & \multicolumn{1}{c|}{author name}                                                      & \begin{tabular}[c]{@{}l@{}}Name of the commenter\end{tabular}                       \\ \cline{2-3} 
\multicolumn{1}{c|}{}                                   & \multicolumn{1}{c|}{author id}                                                        & Id of the commenter                                                                                    \\ \cline{2-3} 
\multicolumn{1}{c|}{}                                   & \multicolumn{1}{c|}{author image}                                                     & \begin{tabular}[c]{@{}l@{}} YouTube profile picture of the\\ commenter\end{tabular}  \\ \cline{2-3} 
\multicolumn{1}{c|}{}                                   & \multicolumn{1}{c|}{comment date}                                                     & \begin{tabular}[c]{@{}l@{}}Date and time the comment \\ was posted\end{tabular}                   \\ \cline{2-3} 
\multicolumn{1}{c|}{}                                   & \multicolumn{1}{c|}{comment text}                                                     & Content of the comment                                                                            \\ \cline{2-3} 
\multicolumn{1}{c|}{}                                   & \multicolumn{1}{c|}{video id}                                                         & \begin{tabular}[c]{@{}l@{}}Id of the video (in which the \\ comment was posted)\end{tabular}      \\ \cline{2-3} 
\multicolumn{1}{c|}{}                                   & \multicolumn{1}{c|}{parent id}                                                        & \begin{tabular}[c]{@{}l@{}}Id of the original comment \\ (if it is a comment reply)\end{tabular}  \\ \cline{2-3} 
\multicolumn{1}{c|}{}                                   & \multicolumn{1}{c|}{like count}                                                       & Number of likes                                                                                   \\ \cline{2-3} 
\multicolumn{1}{c|}{}                                   & \multicolumn{1}{c|}{reply count}                                                      & Number of replies to comment                                                                      \\ \hline
\multicolumn{1}{c|}{\textit{User}}                      & \multicolumn{1}{c|}{\begin{tabular}[c]{@{}c@{}}gender,  age \\ and race\end{tabular}} & features extracted by Face++                                                                     
\end{tabular}

\end{scriptsize}
\end{table}

\subsection{Analysis I: Videos and Channels}
\label{sec:ana1}

\begin{figure*}[!ht]
\centering
\includegraphics[width=1\textwidth]{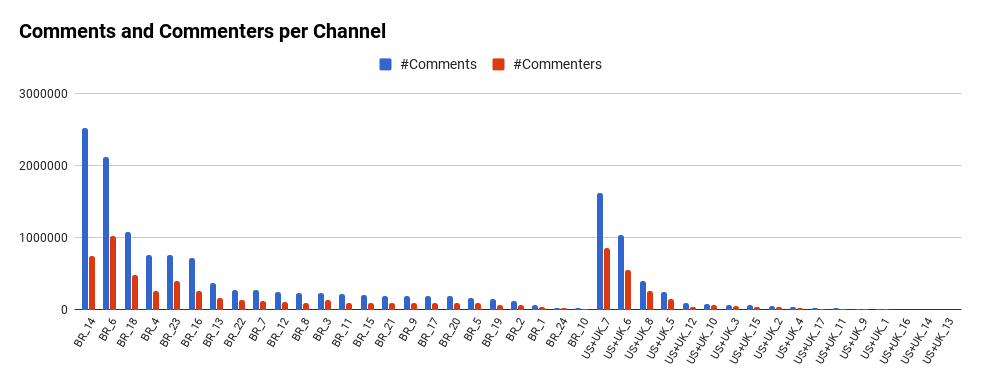}
\caption{Number of commenters and comments per channel.}
\label{fig:usersCommentsChannel}
\end{figure*}

In this section we will present a characterization of the videos present in our dataset. Figure~\ref{fig:usersCommentsChannel} shows the number of comments and commenters - users who have left comments - per channel. In general, Brazilian videos have more commenters and comments, but this may only be a consequence of the selected channels and not a result of the behavior of the audience from the two countries.

It is interesting to notice that the ranking of channels regarding the number of comments and the number of commenters is different. For instance, the channel with more comments in Brazil is BR\_14, while the one with more commenters is BR\_6. This result comes from the fact that the consumers of some channels are more active than others with respect to  interacting (i.e., commenting) with the video.

\begin{figure*}[!ht]
  \begin{subfigure}[t]{0.4\textwidth}
          \centering
          \includegraphics[scale=0.48]{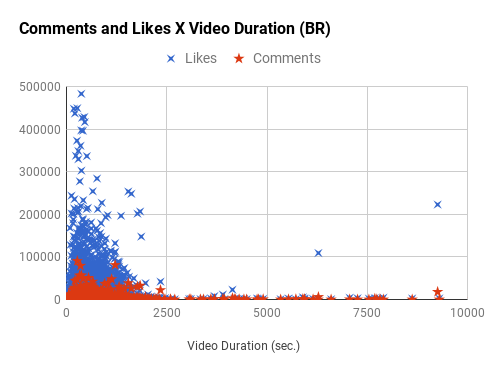}
          \label{fig:CommentsLikesVideoDurationBR}
  \end{subfigure}
	\hspace{5em}
  \begin{subfigure}[t]{0.4\textwidth}
          \centering
          \includegraphics[scale=0.48]{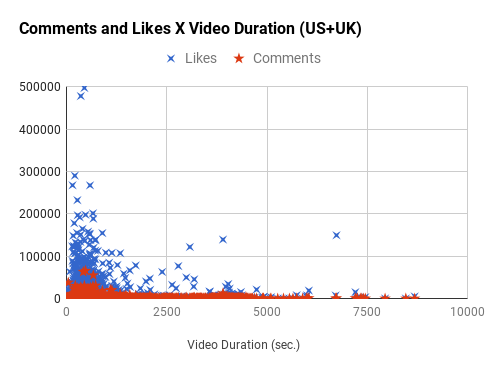}   
          \label{fig:CommentsLikesVideoDurationUS}
  \end{subfigure}
  \caption{Comments and Likes X Video Duration.} 
  \label{fig:CommentsLikesVideoDuration}
\end{figure*}

In Figure~\ref{fig:CommentsLikesVideoDuration} we present the number of comments and likes by video duration in seconds. From the chart we observe that the number of comments and likes does not correlate to the duration of the video, the most popular videos (i.e., videos that receive more likes and comments) are the shortest videos.  

\begin{figure*}[!ht]
\centering
\includegraphics[scale=0.6]{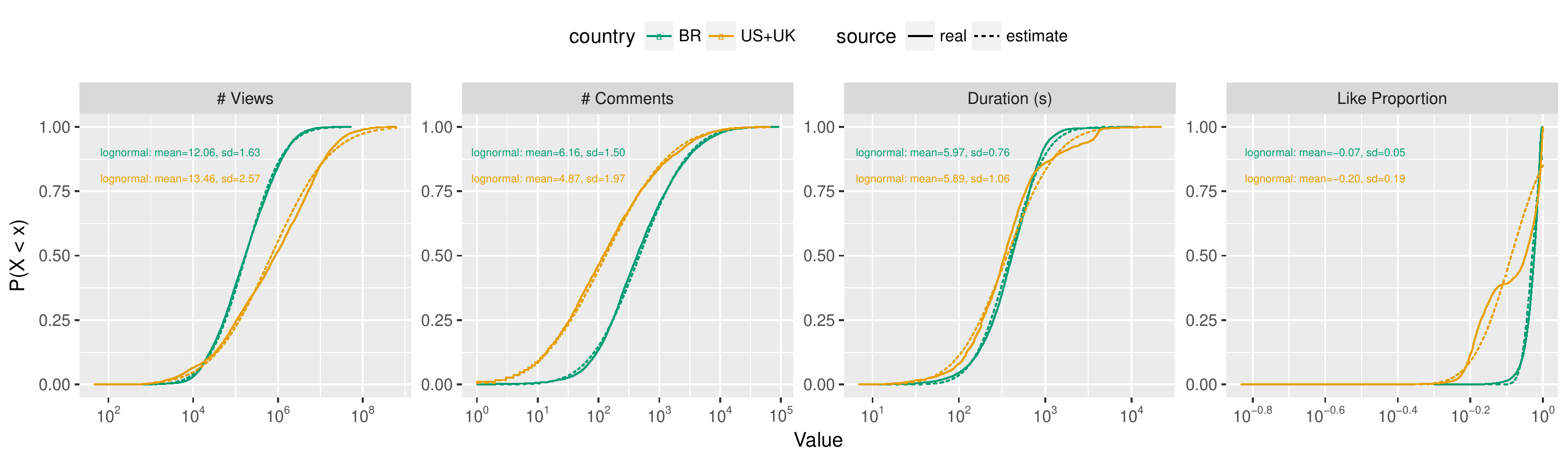}
\caption{Distribution of Video Statistics} 
\label{fig:video_stats}
\end{figure*}

Figure~\ref{fig:video_stats} presents the cumulative distributions of four video metrics (number of visualizations, number of comments, duration in seconds and proportion of likes), comparing between the countries. We fitted the metrics to a Lognormal distribution, which probability density function is given by the formula $$f(x) = \frac 1 x \cdot \frac 1 {\sigma\sqrt{2\pi\,}} \exp\left( -\frac{(\ln x-\mu)^2}{2\sigma^2} \right)$$. We estimate the parameters $\mu$ (mean) and $\sigma$ (sd) using the maximum likelihood estimation technique. We present the corresponding function of distribution and the estimated parameters in the plots. 

As we observe, the shape of the curves are similar between both datasets, although presenting different values. For instance, US and UK videos have more visualizations, while Brazil videos have more comments and a higher proportion of likes.

\subsection{Analysis II: Advertising}
\label{sec:ana2}

According to Westenberg~\cite{westenberg2016influence}, YouTubers are viewed as authentic by their audience, when reviewing a product or brand. Followers believe that Youtubers' recommendations are honest.  In order to look more honest and transparent to their followers, Youtubers  label their promoted  videos with special hashtags, meaning the content, product or brand is sponsored. Thus, we take advantaged of the presence of hastags to identify the commercial nature of a video.  Videos may contain explicit  or implicit advertising. The former  involves  direct sales messages to a target audience. Implicit advertising, on the other hand, works best when businesses want to associate their brand or products with a psychological or symbolic element.  We argue that if a video mentions products or brands, it potentially  has advertising messages. To verify whether a video has advertisement, we employ the methodology of video classification explained in Section~\ref{sec:method_video_classification}. We were able to classify 219 out of 1,055 tags for Brazil, and 249 out of 1,010 tags for channels in English. In total, 6,017 videos in Brazil were classified, and 4,109 videos in English.


\begin{figure}[!ht]
\centering
\includegraphics[scale=0.5]{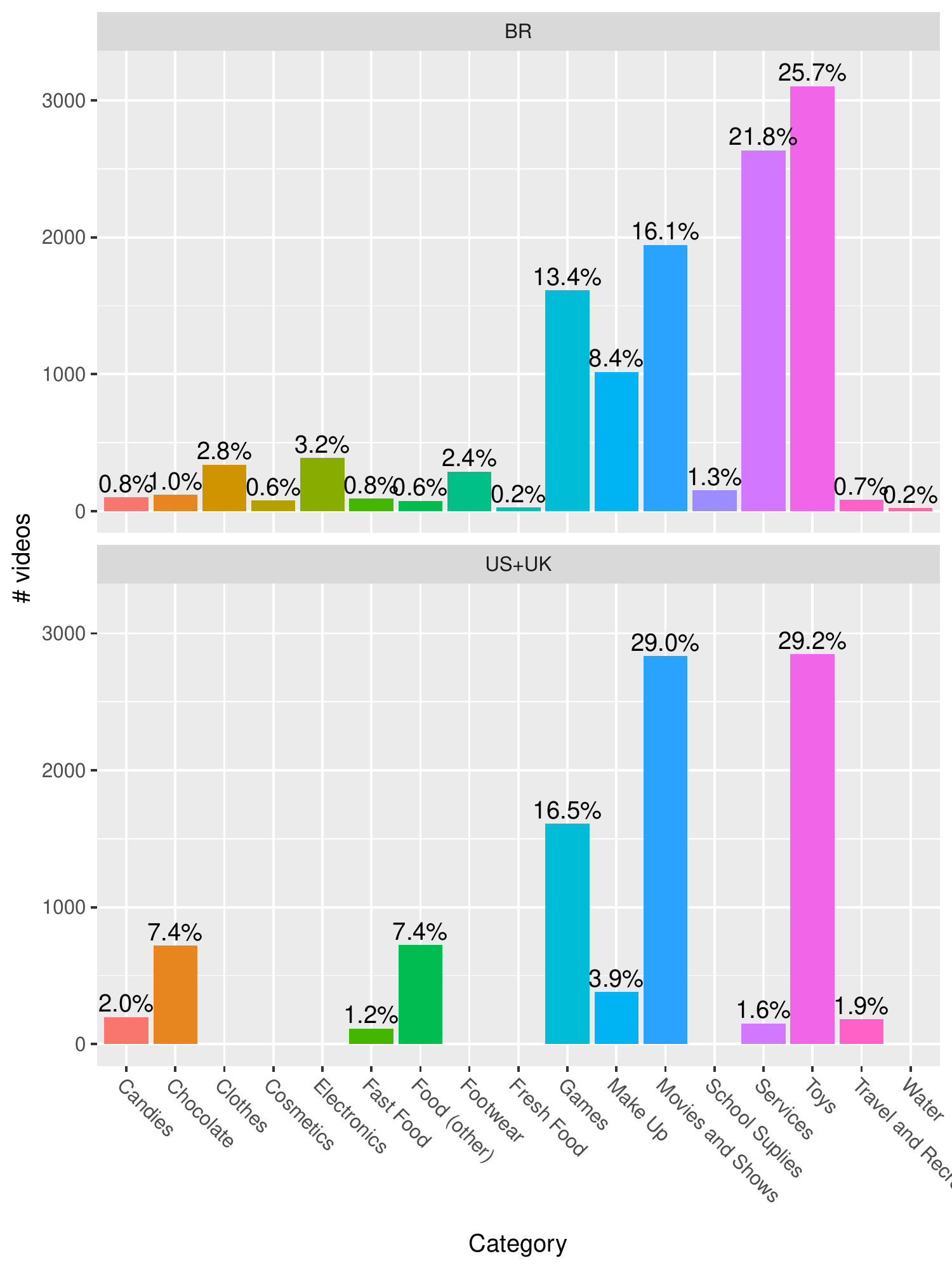}
\caption{Frequency of Video Categories by Country} 
\label{fig:video_categories_general}
\end{figure}

Figure~\ref{fig:video_categories_general} shows the distribution of the video categories for both Brazil and English channels. The categories "Toys", "Movies and Shows" and "Games" are very popular in both countries, while "Services" is popular only in Brazil.

\begin{figure*}[!ht]
\centering
\includegraphics[scale=0.5]{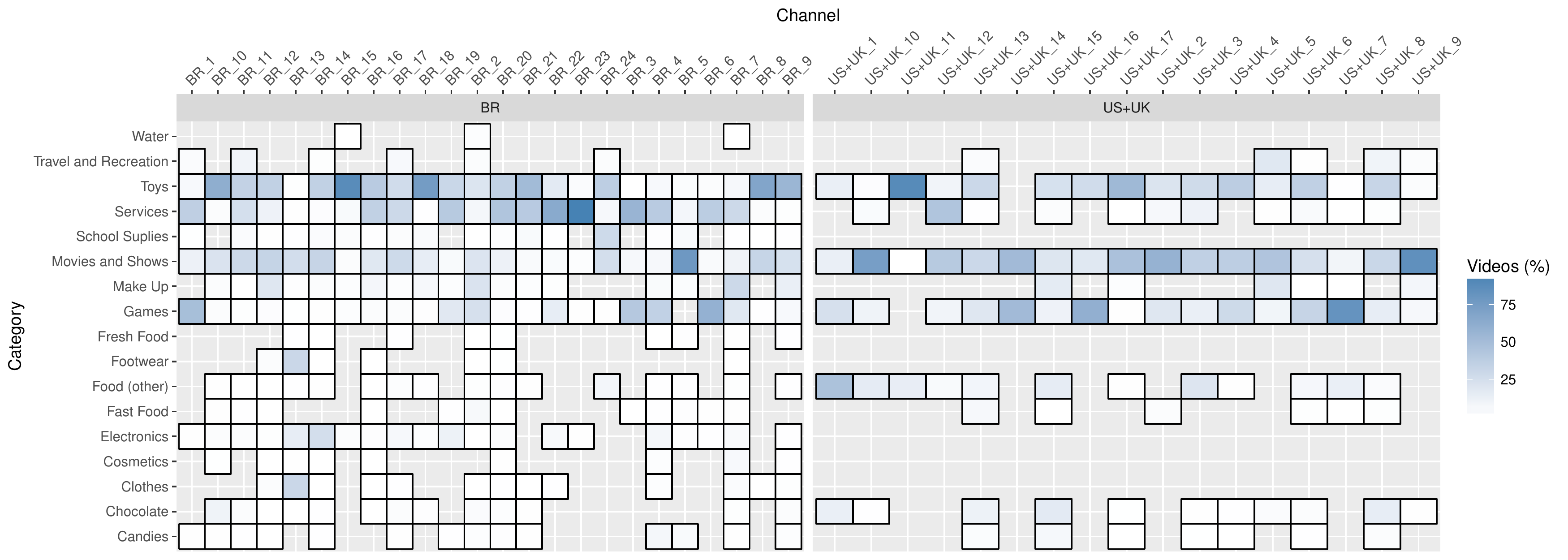}
\caption{Frequency of Video Categories by Channel} 
\label{fig:video_categories_channel}
\end{figure*}

Figure~\ref{fig:video_categories_channel} presents the distribution of categories among all the 41 channels. We observe that there are some channels specialized in a certain kind of content. For instance, channels BR\_15 and US+UK\_11 have a high proportion of videos about Toys. Channels BR\_6 and US+UK\_7 are mostly about Games.

\subsection{Analysis III: Audience}
\label{sec:ana3}

In this section our analysis focuses on YouTube audience. Considering only videos that have some kind of advertisement (i.e. those we were able to classify into one of the categories). The average number of comments per user is 3.19, for Brazilian videos, and 1.74, for English channels.

\begin{figure*}[!ht]
\centering
\includegraphics[scale=0.7]{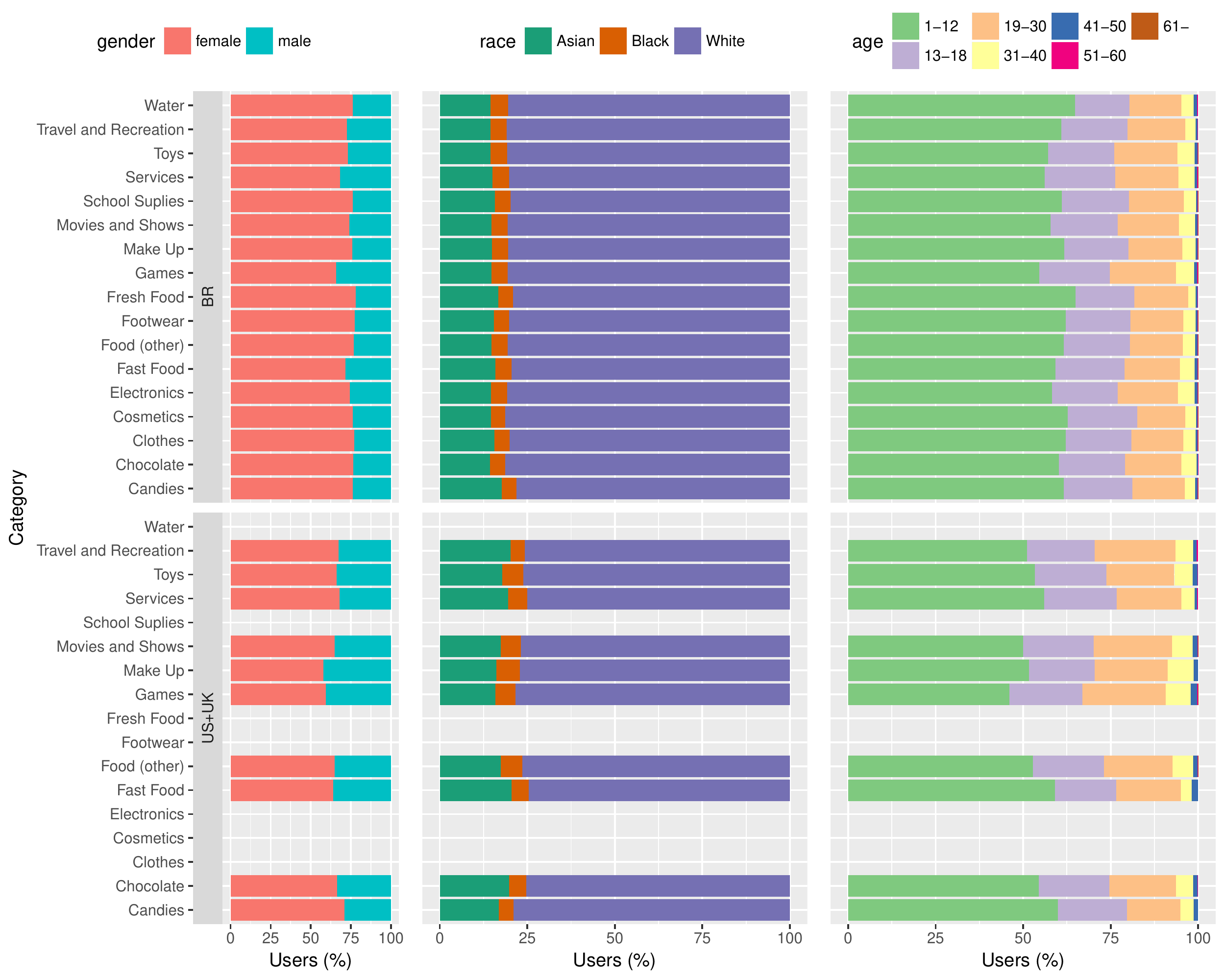}
\caption{Demographics of Audience by Category} 
\label{fig:user_profile_category}
\end{figure*}

\begin{figure*}[!h]
\centering
\includegraphics[scale=0.7]{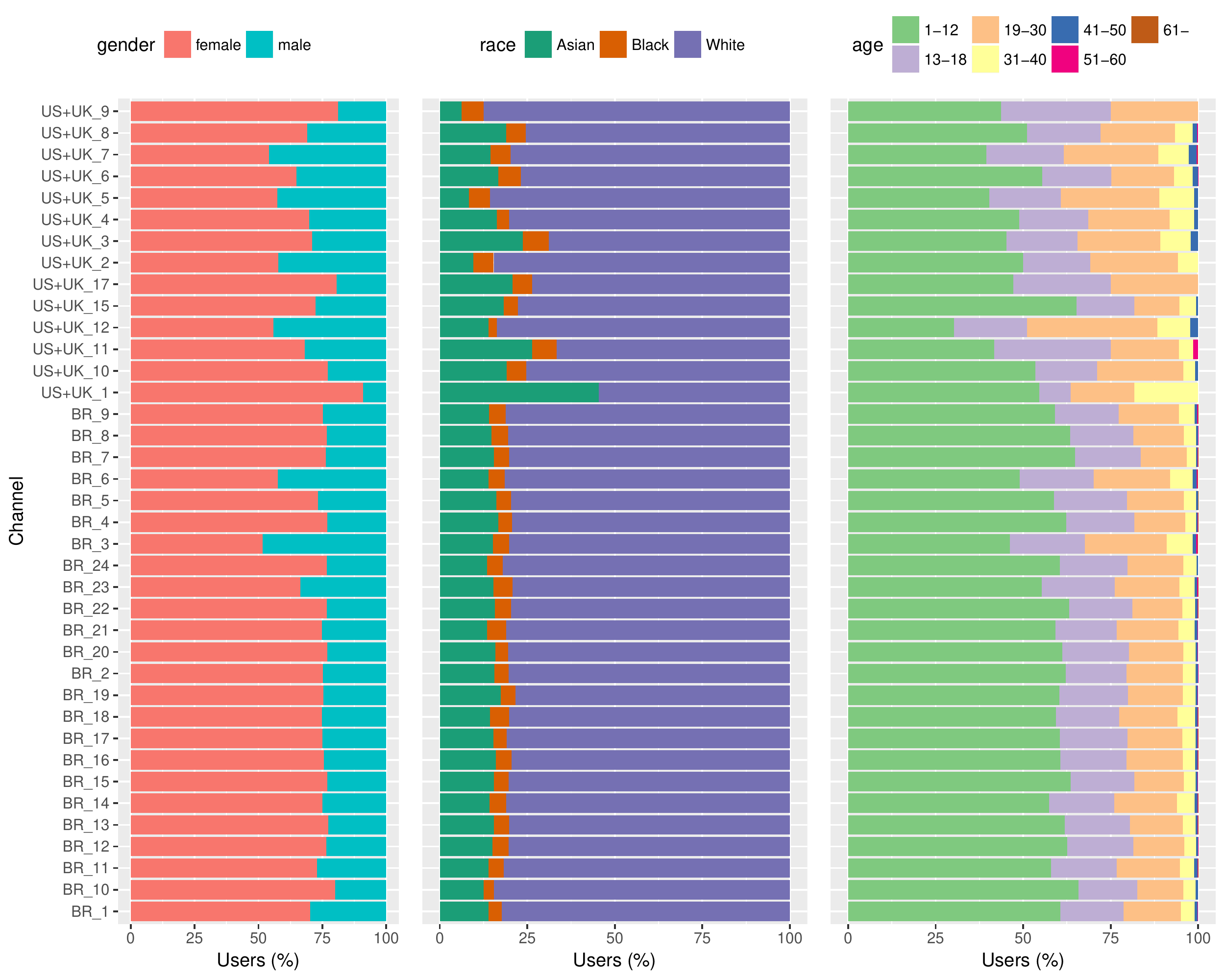}
\caption{Demographics of Audience by Channel} 
\label{fig:user_profile_channel}
\end{figure*}

\begin{figure}[!h]
\centering
\includegraphics[scale=0.7]{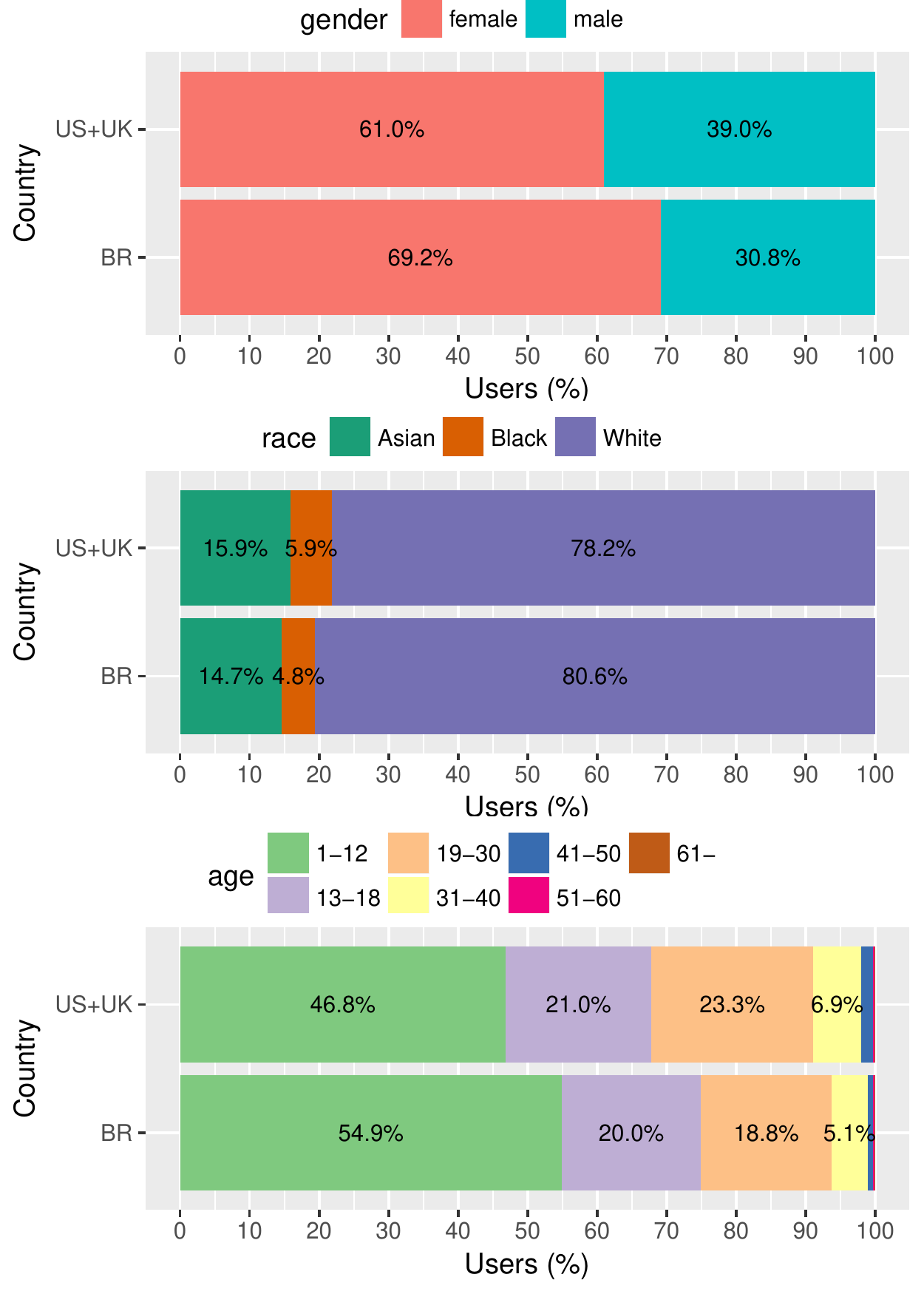}
\caption{Demographics of Audience by Country} 
\label{fig:user_profile_country}
\end{figure}

\begin{figure}[!h]
\centering
\includegraphics[scale=0.5]{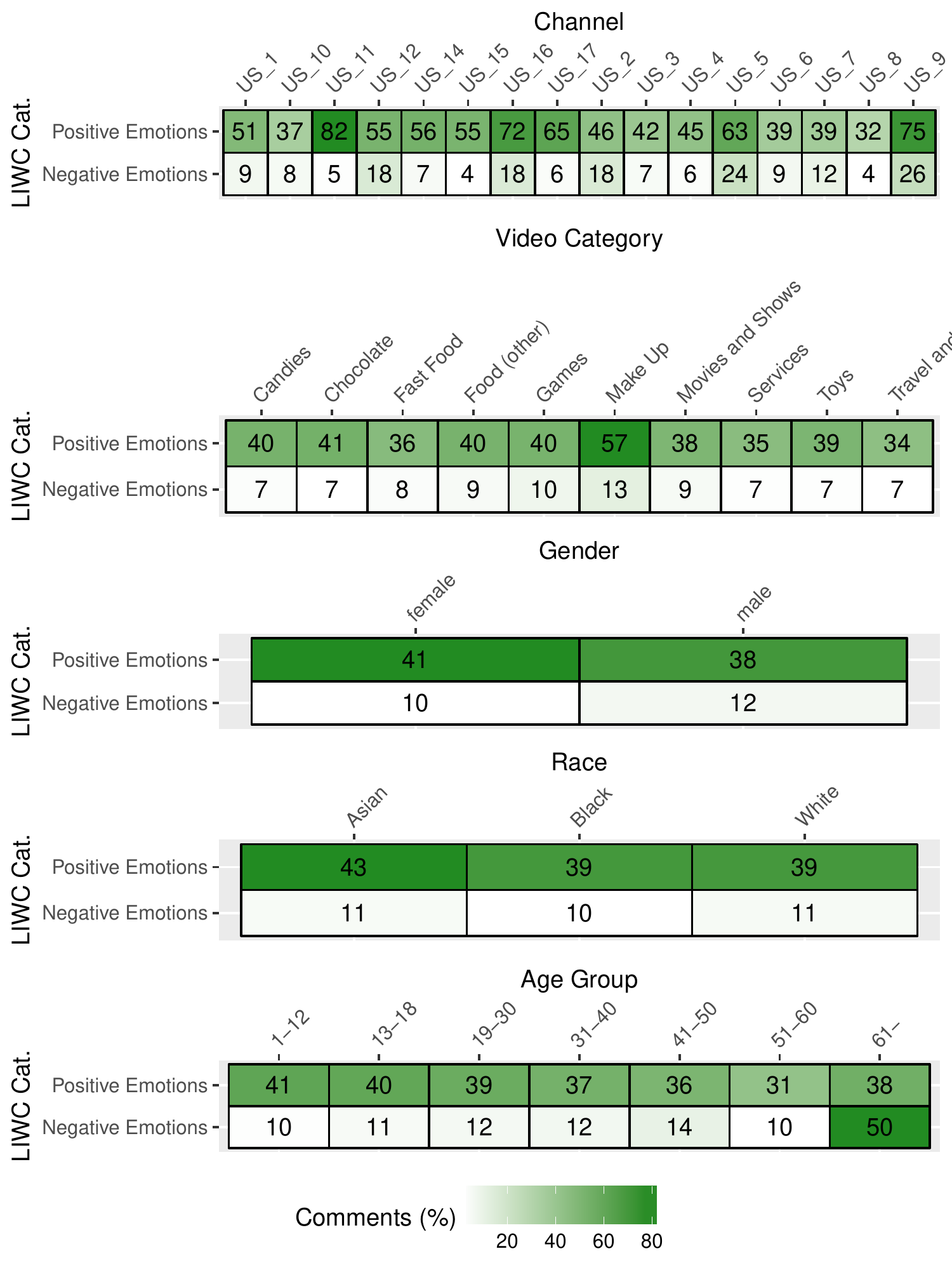}
\caption{Percentage of Comments Containing words of Positive and Negative Emotion}
\label{fig:comments_liwc_emotions}
\end{figure}

First, we show gender, race and age distributions of commenters in Figure~\ref{fig:user_profile_country} - considering only videos with advertisement. The difference between men and women is smaller for users in English channels, but the proportion of women is higher in both countries. In the race distribution (middle) we observe a similar distribution for both countries showing that the proportion of white people commenting on videos is higher than the proportion of Asian and black people. On the bottom chart we observe that the user age distribution for both countries present similar behavior, the only difference is that in Brazilian videos the second age group who most commented on youtuber are children and teenagers between 13 and 18 years and, for the English channels, they are young people from 19 to 30 years. We choose these age intervals to distinguish from children under 12 years old, who should not have a YouTube account, and children and teenager audience between the ages of 13 and 18 who may have an account under parental supervision.

Figure~\ref{fig:user_profile_channel} shows the demographics with respect to gender, race and age groups for each channel. We observe that the profile of the audience might be very different. For instance, 90\% of channel US+UK\_1's audience is female, while this is only 52\% for channel BR\_3. Regarding age, some channels have a child audience of nearly 70\%, such as channels US+UK\_15 and BR\_10.

Figure~\ref{fig:user_profile_category} shows the demographics for each category, for both countries. We observe that the Games category is the one with higher proportion of male audience, although still having more women comparatively. Also, the audience for Games is older, presenting lower frequencies of children.

Brazilian legal framework considers publicity aimed at children as abusive and, therefore, illegal. The legal definition of children includes any person under 12 years, and the Consumer Code, which specifies a set of abusive conducts, lists as one of them the act of directly approach children with publicity of products or services, considering they haven’t reached a certain degree of bio-physical development which is necessary to identify and understand the marketing discourse and, therefore, is legally regarded as vulnerable. Brazilian law regarding Internet - basically the Internet Civil Rights Framework, or Marco Civil da Internet - basically follows this same rationale and recognizes the need for special information and education about children's access to Internet. There is no general data protection framework enacted in Brazil which could impact children’s privacy.

The collection and use of Children’s personal data is subject to the standards of COPPA (Children's Online Privacy Protection Act of 1998), which dictates that no data regarding persons under 13 years can be collected without their parents or caregivers  explicit consent. COPPA also includes a series of obligations for site owners, makes it mandatory for a website to include in its privacy policy a set of rules and warranties for the its usage by children, and also clarifies how the consent from the parent or responsible has to be collected.

\subsubsection{Semantics}

In this analysis we want to measure how the video was evaluated by the viewers. We use the text of the comments as a proxy for the perception of the audience, looking into the semantics it contains. We focus on only two categories of LIWC: Positive Emotions and Negative Emotions. Since the LIWC is available only for the English language, we inspect only comments from the U.S. channels.

Figure~\ref{fig:comments_liwc_emotions} presents the percentage of the comments that contain words related to positive emotions or negative emotions, according to LIWC. We aggregate the comments by channel, video category, gender and age group. The predominance of positive emotions is notorious, indicating that the videos are, in general, well evaluated by the public. Interestingly, some channels have a higher proportion of positive words than others, such as US+UK\_11 and US+UK\_9. Regarding the video categories, we observe that videos with make up are more positive than the others. Looking into the social groups, there are no huge differences, although we observe an indication that the use of positive words seems to decrease as the audience get older.

\section{Conclusions}
\label{conclusion}

Google has some clear age policies for its  products\footnote{support.google.com/accounts/answer/1350409?hl=en}.  The minimum age requirements to own a Google account in the United States  is  13 or older (i.e., except for Google Accounts created in Family Link for kids under 13), 14 or older in Spain and South Korea, 16 or older in Netherlands and 13 or older in all other countries. Some services have specific rules, such as YouTube that specifies that  age-restricted video  should be watched only by users who are 18 or older.  In Brazil, however, the use of YouTube itself is restricted to those over 18, according to the terms of service of the platform that is clear in stating that   the YouTube website is not designed for young people under 18 years\footnote{www.youtube.com/static?gl=BR\&template=terms\&hl=pt}.

Among outcomes of this paper we could also mention that it can lead to a discussion about Google’s politics on the age limit if it is confirmed the active presence of under 13 on YouTube. Also, data gathered and analyzed about the profile of users under 13 may be used in future research about (i) the presence of racial and gender bias, (ii) the means publicity approaches children on YouTube and (iii) the way private data from children is collected and commercialized in digital media.

Out of the data analyzed, we believe the major impact may result from the identification and characterization of children actively using YouTube. Even if some usage of under-18 is generally considered as fair due to parents’ or legal responsible consent and supervision, the fact is that if children are actually using the platform they can be exposed to different challenges, as advertising, inappropriate content, privacy issues and crimes in the digital world, which raise concerns about compliance with regulations in several countries.

Other questions addressed by this work could be investigated in greater detail to highlight possible nuances not captured by the experiments done. For example, an evaluation of how exactly Face++ accuracy is impacted by the particularities of the pictures in the user profile could help to know whether there are adjustments to be made in this respect. A detailed analysis of usage patterns and spread of YouTube channels  across countries may reveal how local differences affect the overall temporal dynamics found. Analysis of the influence of geographic and cultural location on the user behavior would be interesting for promoting educational and healthy food videos among children. Considering that many videos   blur the boundaries between entertainment and advertising~\cite{dredge2015why}, another possible direction would be to dig further into the transcripts of the videos to analyze the texts and characterize the different types of advertising that are exhibited to children.

\section*{Acknowledgements}

This work was partially supported by CNPq, CAPES, FA\-PE\-MIG, and the projects InWeb, MASWEB, and INCT-Cyber.
%
%
%
%
%
%
%
%
%
%
%
%
%
%
%
%
%
%

\bibliographystyle{abbrv}
\bibliography{sigproc}



\end{document}